\documentclass[prl,twocolumn,showpacs]{revtex4}
\usepackage{dcolumn}
\usepackage{graphicx}
\usepackage{amsmath,epsf}
\usepackage{amssymb}
\begin{document}
\title{Efficient Selfconsistent Calculations of 
Multiband Superconductivity in UPd$_2$Al$_3$}
\author{P. M. Oppeneer$^1$ and G. Varelogiannis$^{2}$}
\affiliation{
$^1$Institute of Solid State and Materials Research, P.O. Box 270016,
D-01171 Dresden, Germany\\
$^2$Department of Physics,
National Technical University of Athens, GR-15780 Athens, Greece}
\date{\today}
\begin{abstract}
An efficient physically motivated computational approach 
to multiband superconductivity is introduced
and applied to the study of the gap symmetry
in a heavy-fermion, UPd$_2$Al$_3$.
Using realistic pairing potentials and accurate energy bands that are
computed within density functional theory, 
self-consistent calculations demonstrate
that the only accessible superconducting
gap with nodes exhibits $d$-wave symmetry in  
the $A_{1g}$ representation of the $D_{6h}$ point group.
Our results suggest that in a superconductor with                    
gap nodes the prevailing gap symmetry is dictated by
the constraint that nodes must be as far as possible
from high-density areas.
\end{abstract}
\pacs{PACS numbers: 74.25.Jb, 74.20.-z}
\maketitle

Superconductivity (SC)
in heavy-fermion (HF) materials exhibits a
fascinating complexity of phenomena \cite{Steglich}.
The nature of the SC state is mostly anticipated to be
unconventional, i.e., an additional symmetry is broken in the SC state,
and the order parameter is therefore not the conventional spin singlet
$s$-wave type
\cite{SigristUeda,Annett}.
The identification of the order parameter symmetry is 
an essential
issue in current studies of unconventional SC.
A common approach to address the symmetry of the order parameter is through
investigations of the SC gap, which is expected to have the
same symmetry as the order parameter.
While progress has been made for
high-$T_c$ superconductors (HTSC),
the identification of the gap symmetry remains
an open question for most HF materials.
The eventual identification of the gap
symmetry in a HF compound
is, however, not sufficient to identify the pairing mechanism.
Additional elements are necessary and
in the case of HF materials, it was suggested that spin fluctuations~-
either in the itinerant 
\cite{mathur98,saxena00}
or localized limit as magnetic excitons 
\cite{SatoNat,thalmeier02}~-
may replace
the phonons as mediators of the pairing.

In order to identify the gap symmetry and clarify 
its relationship to 
the pairing mechanism,
selfconsistent solutions of the SC gap equations with
physical momentum dependent pairing potentials
and accurate band structures are unavoidable ({\it cf.}
\cite{temmerman96,mazin99}).
Unfortunately,
in the case of HF's the situation is rather
complex. In addition to the three-dimensionality,
we must deal with several anisotropic bands
that cross the Fermi level ($E_F$) producing numerous
highly anisotropic Fermi surface (FS) sheets.
As a consequence, no
realistic selfconsistent calculations
of SC have been achieved so far in HF
materials.

Here we introduce a            
computational approach to multiband SC in a HF material,
by directly combining relativistic band-structure calculations
with a physically motivated
procedure to solve efficiently and accurately
the multiband BCS gap equation
selfconsistently.
The efficiency of our procedure allows a systematic study of
the influence of the momentum structure of the pairing on the
resulting gap symmetry in an anisotropic multiband system.
We focus on UPd$_2$Al$_3$ which is a
fascinating HF superconductor having
a moderately large specific heat
coefficient $\gamma = 140$\,mJ/mol\,K$^2$, and a
relatively high critical temperature
$T_c = 2$\,K   
\cite{geibel91}.
It orders antiferromagnetically
below $T_N = 14.3$\,K with an ordered moment of 0.85\,$\mu_B$/U-atom
\cite{krimmel92},
which is large compared to the moments of other HF superconductors
as, e.g., UPt$_3$, which has a moment of only 0.03\,$\mu_B$
\cite{joynt02}.
A further anomalous feature is that
the antiferromagnetic (AFM) order coexists with SC below 2\,K.
Coexistence of magnetism and SC was observed for other
materials, e.g., containing $4f$ elements,
but in those cases the magnetism is due to the localized $4f$ electrons
far from $E_F$,             
whereas the SC is carried by itinerant electrons at $E_F$.           
Conversely, for UPd$_2$Al$_3$ most of the recent 
studies reveal that
the SC, magnetic order, as well as HF behavior {\it all} involve
the uranium $5f$ states \cite{knopfle96,bernhoeft00}.

Previous investigations revealed that UPd$_2$Al$_3$ can be classified 
as a spin singlet,
BSC-type superconductor \cite{amato92,kyougaku93,huth00}.
The appropriate starting point is therefore the 
coupled multiband BCS gap equation \cite{suhl59}, for $T\rightarrow 0$:
\begin{equation}
\Delta_n (\boldsymbol{k}) = - \sum\limits_{n^{\prime}}
\sum\limits_{\boldsymbol{k}^{\prime}}
\frac{ V_{nn^{\prime}}(\boldsymbol{k},\boldsymbol{k}^{\prime}) \Delta_{n^{\prime}}
(\boldsymbol{k}^{\prime})}
{2 \left[{ E^2_{n^{\prime}{\boldsymbol{k}^{\prime}}} + |\Delta_{n^{\prime}}
(\boldsymbol{k}^{\prime})|^2 }\right]^{1/2}} ,
\end{equation}
where $\Delta_n$ is the gap of band $n$, and $E_n$ is the corresponding 
singleparticle
energy relative to $E_F$. $V_{nn^{\prime}}$ is
the pairing potential. The coupled multiband gap equation is at present 
too complicated to be solved numerically
on the real FS of the material and with
physical momentum-dependent pairing potentials. 
However it is a common
experimental practice not to consider a gap symmetry defined separately
for each band, but to work with a ``global'' gap symmetry valid for all bands,
i.e., defined in the whole Brillouin zone (BZ)
(see, e.g., \cite{joynt02,jourdan99,hessert97}).
Our goal is to compute selfconsistently such global gap symmetry.
To achieve this goal, 
one could proceed as follows:
solve for each individual band
the singleband BCS gap equation, 
and subsequently, combine and interpolate these singleband gaps to a generalized
gap $\Delta (\boldsymbol{k})$ defined in the whole BZ.
This is precisely what we do by solving as an {\it ansatz} a modified gap equation
\begin{equation}
\Delta (\boldsymbol{k}) = - 
\sum\limits_{\boldsymbol{k}^{\prime}}
\frac{ V (\boldsymbol{k},\boldsymbol{k}^{\prime}) \Delta
(\boldsymbol{k}^{\prime})}
{2 \left[{ G({{\boldsymbol{k}^{\prime}}}) + |\Delta
(\boldsymbol{k}^{\prime})|^2 }\right]^{1/2}} \, ,
\end{equation}
with $G(\boldsymbol{k}) \equiv (1/E^2_{n_1 \boldsymbol{k}} + 
1/E^2_{n_2 \boldsymbol{k}}
+ \cdots )^{-1}$. 
On the FS sheet of band $n$,  
$G_{\boldsymbol{k}}$ is equal to  $E^2_{n\boldsymbol{k}}$, and thus
$\Delta (\boldsymbol{k}) = \Delta_n (\boldsymbol{k})$.
Note that intraband as well as interband processes are taken into consideration.
A simplification specific to UPd$_2$Al$_3$ has been implemented
in Eq. (2) which could,
however, be relaxed. Since all bands 
at $E_F$ have mainly $5f$ character \cite{knopfle96}, there is 
no need to discriminate different band dependencies in the
pairing potential, which we thus assume to be identical for all $n$.

The energy dispersions $E_n (\boldsymbol{k})$
were computed within the framework of density functional
theory in the local spin-density approximation (LSDA).
Kn\"opfle {\it et al}.
\cite{knopfle96} demonstrated a very close agreement between
the measured \cite{inada94} and calculated in this framework
de Haas-van Alphen frequencies.
Consequently, the FS of UPd$_2$Al$_3$ is accurately described by the
LSDA energy dispersions. Since the relevant physics for SC happens in the
vicinity of the FS, the LSDA energies provide the appropriate input for
our investigation of the gap symmetry.
We computed the LSDA energy bands of UPd$_2$Al$_3$ using the relativistic 
augmented-spherical-wave method
\cite{williams79}.
The resulting
band structure of UPd$_2$Al$_3$ along
characteristic symmetry lines is shown in Fig.~1         
for the paramagnetic and AFM state.
The FS resulting from our AFM bands
consists of four types of sheets,
and is practically identical to that computed previously \cite{knopfle96},
as is also the ordered AFM moment of 0.81~$\mu_B$, which  
agrees with the experimental moment of 0.85~$\mu_B$
\cite{krimmel92}.

For the pairing potential we consider here
two categories of pairing kernels
that are generally accepted and were shown to produce unconventional
order parameters.
The small-${q}$ pairing potential,
adopts at small wavevectors $\boldsymbol{k} - \boldsymbol{k}^{\prime}$
the following form in momentum space
$
V(\boldsymbol{k}, \boldsymbol{k}^{\prime})=$ 
$\frac{-V} { \boldsymbol{q}_c^2+(\boldsymbol{k}-\boldsymbol{k}^{\prime})^2}+
$ $\mu^*(\boldsymbol{k}-\boldsymbol{k}^{\prime})$.
This kernel is characterized by a smooth momentum cut-off $\boldsymbol{q}_c$
which selects the small-${q}$ processes in the
attractive part
 \cite{smallq1,Abrikosov,smallq1a,Weger,Leggett}.
At larger wavevectors the repulsive
Coulomb pseudopotential $ \mu^*({\boldsymbol{k}-\boldsymbol{k}^{\prime}})$ 
may prevail.
This type of attractive interaction
may be due to {\it phonons} if
the system is close to instabilities and screening with short range
Hubbard-like Coulomb terms is involved
\cite{smallq1a}.
It has been claimed that this type of kernel 
may account for unconventional SC in HTSC       
\cite{smallq1,Abrikosov,smallq1a,Weger,Leggett} and
in other materials, including HF's
\cite{smallq1a}.
Agterberg {\it et al}. 
\cite{Agterberg}
have recently suggested for
HF superconductors 
a multipocket FS with an attractive
intrapocket and repulsive interpocket potential.
Such situation results naturally from a potential
dominated by small-${q}$ attractive pairing as the one
considered here,
if the FS contains multipockets
separated in momentum space. 
We also considered the case of pairing mediated by 
{\it spin fluctuations} 
using a phenomenological
Millis-Monien-Pines pairing potential  
\cite{millis90}
which has the form
$V(\boldsymbol{k}, \boldsymbol{k}^{\prime} )=
\frac{V}{{\boldsymbol{q}}_c^2+({\boldsymbol{k}}-{\boldsymbol{k}^{\prime}}
-{\boldsymbol{ Q}})^2}$
where ${\boldsymbol{ Q}}=\pm(0,0,\pi/2c)$ in UPd$_2$Al$_3$ 
($c$ is the $c$ axis      
lattice constant).
The $\boldsymbol{Q}$ value was verified by neutron scattering experiments
\cite{bernhoeft98}.
Pairing through spin fluctuations has been suggested for
virtually all unconventional SC's, including UPd$_2$Al$_3$
\cite{bernhoeft00,huth00}.

To obtain a selfconsistent solution of our modified gap equation 
with such momentum dependent kernels
and the original LSDA bands
we have used a Fast-Fourier-Transform technique.
The problem is solved iteratively on a symmetric             
part of the BZ
that we discretize with a
$128\times 128 \times 128$ momentum grid. 
For each point of our 3-D grid we have
our LSDA energy bands as an input.
Within our procedure the momentum space problem is fully
resolved numerically without any simplification or          
bias on the resulting gap symmetry. 

The accessible even parity gap states can posses $s$- or $d$-wave
symmetry. 
As UPd$_2$Al$_3$ and thus our LSDA bands obey the hexagonal
$D_{6h}$ point group symmetry, the accessible gap states  
of $d$-wave symmetry that have nodes 
should transform according to
the irreducible representations of the
$D_{6h}$ group 
\cite{Yip} 
shown schematically in Fig.~2.
In our calculations we used the small-$q$ phonon
pairing potential
with various parameter choices for both the paramagnetic and AFM
energy bands 
and the spin-fluctuations pairing potential only for the AFM bands.
An initial gap configuration is chosen randomly and the system
converges within the iteration cycle towards the most favorable solution.
Also, we adopted as initial gap each of the representations in Fig.~2
and tried if the system would converge towards the chosen configuration.
Depending on the parameters in pairing kernel 
($\boldsymbol{q}_c$, $V$, $\mu^*$), we can achieve as 
selfconsistent solution either a gap having
$s$-wave or $d$-wave symmetry. 
Quite unexpectedly, in the latter case we find that
{\it the only accessible 
gap state with nodes in UPd$_2$Al$_3$
belongs to the irreducible representation $A_{1g}$}. No other representation
possessing nodes was accessible for all the parameters in
both types of pairing kernels.

We show in Fig.~3 some examples of our
selfconsistently computed gap functions
in the plane $\rm H - A - H $, $\rm H - K - H$
which contains
the $z$ axis (the $ {\rm A} - \Gamma - {\rm A}$ axis) and in the
plane obtained by a $\pi/6$ or $\pi/2$ rotation around the $z$ axis
(the plane $ {\rm L - A - L}$, $\rm  L - M - L$). All selfconsistent 
solutions shown in Fig.~3 have
two lines of nodes perpendicular to the ${\rm A} - \Gamma - 
{\rm A}$ axis, thus
belonging in the $A_{1g}$ representation.
To study the influence of the coexisting AFM
order on SC we have made calculations using both paramagnetic and
AFM bands.
In both paramagnetic and AFM cases, the only solutions with
gap nodes are of $A_{1g}$ type.
We note that our results
with the spin-fluctuations kernel
confirm the model analyses performed by Huth {\it et al}. \cite{huth00}
and Bernhoeft \cite{bernhoeft00}. In particular, Bernhoeft predicted
for SC mediated by spin fluctuations 
in UPd$_2$Al$_3$ a characteristic symmetry property of the global gap, {\it viz.}
$\Delta(\boldsymbol{k}) = - \Delta(\boldsymbol{k} + \boldsymbol{Q})$,
which, in its simplest form, would be fulfilled by 
a gap having the
functional dependence $\Delta (\boldsymbol{k}) \propto \cos (k_zc)$. 
Our gap computed selfconsistently with the spin-fluctuation pairing
kernel does obey this symmetry property, and the shape of the
gap does correspond closely to $\cos (k_zc)$ (see Fig.~3).
In addition, we find here that small-$q$ phonon pairing may lead to the
same $d$-wave $A_{1g}$ state, demonstrating that there is not
a one-to-one correspondence between the gap symmetry and a
specific microscopic mechanism, although    
experiments suggest that spin degrees of freedom 
are likely involved in the pairing \cite{bernhoeft98,metoki98,SatoNat}.

There is overwhelming experimental evidence that
indeed UPd$_2$Al$_3$
has a gap with the
$A_{1g}$ node structure. 
Recent tunnel measurements along the $z$ axis showed the absence of a
node in this direction \cite{jourdan99}. However,
the presence of nodes is definitely established for UPd$_2$Al$_3$, and
since $A_{1g}$ is
the only even parity gap with nodes which is nodeless
along the $z$ direction (see Fig.~2), 
it was deduced {\it a posteriori} that the
gap symmetry must be $A_{1g}$ type
\cite{jourdan99}.
In addition, the observation \cite{metoki98,bernhoeft98} of
a spin-fluctuations peak below $T_c$ at $(0,0,\pi/2c)$
would be forbidden by
the involved
coherence factors unless the gap changes
sign along the $z$ axis as in
the $A_{1g}$ representation.
Finally, measurements of the angular dependence
of the critical field indicate two lines of
nodes perpendicular to the $z$ axis 
\cite{hessert97} 
again suggesting
the $A_{1g}$ representation. UPd$_2$Al$_3$
is one of the few  HF superconductor for which
such a clear picture of the nodal structure of the gap is obtained
from the experiments. We consider the agreement of our results
with the experiments as strong support of our calculations.

The surprising robustness of the $A_{1g}$ solution
results from the following general rule:
If a system must choose a gap symmetry with nodes
because of the repulsive effective interaction
at large wavevectors (short distances)
it chooses {\it the representation that
has the minimum number of nodes as far as possible from the
high-density areas}.
In fact, the system must  {\it maximize} the condensation
free energy and this is obtained when there is a gap
in the high-density areas and the node (gapless)
areas are kept minimal.
Examining the paramagnetic bands of UPd$_2$Al$_3$ (Fig.~1b)
we see that the high-density areas near
the FS due to saddle points in the band dispersions
are found essentially near the A point and
near the H point. The $A_{1g}$ representation
prevails because it is the
only one {\it without a node near the A point}.
In the case of the AFM bands (Fig.~1c), the saddle points
of the bands near the FS are found
essentially in the $z=0$ plane (near the $\Gamma$ point,
along the $\Gamma-\rm K$, $\rm K-M$ and $\rm M-\Gamma$ 
symmetry lines)
and near the $\Gamma$ point along $\Gamma - \rm A^{\prime}$
 ($\rm A^{\prime} = A/2$).
The high-density areas in the $z=0$ plane exclude
the $B_{1g}$, $E_{1g}$ and $B_{2g}$ representations which
have a line node in the $z=0$ plane.
From the remaining $A_{1g}$, $E_{2g}$ and $A_{2g}$
representations, $A_{1g}$ is the only one without
a line of nodes parallel to the $\rm A-\Gamma-A$ axis
which would cross the high-density areas at
$\Gamma$ and near $\Gamma$ along $\rm \Gamma-A^{\prime}$.
Also, due to the AFM symmetry a high density occurs again at
$\rm A$, as $\rm A$  is equivalent to $\Gamma$.
The node perpendicular to the $\rm \Gamma-A$ path
in the $A_{1g}$ representation is 
close to $\rm A^{\prime}$ and therefore                
does not cross the high-density areas near $\Gamma$ and 
$\rm A$.
Note that $A_{2g}$, $B_{1g}$ and $B_{2g}$ are
also handicapped by the fact that they have more
nodal areas.

To stress the generality of this argument we reconsidered the
case of cuprate HTSC.
Saddle points produce
high-density areas in the vicinity of $(0,\pi/a)$ and symmetry related
areas ($a$ is the basal-plane lattice constant). 
The $d_{x^2-y^2}$ gap symmetry corresponds to nodes along the
$(\pi/a,\pi/a)$ direction as far as possible from the
high-density $(0,\pm\pi/a)$ and $(\pm\pi/a,0)$ areas.
Using both types of pairing kernels as adopted 
for UPd$_2$Al$_3$ (now $\boldsymbol{Q}=(\pi/a,\pi/a)$)
we were never able to obtain
a $d_{xy}$ solution for which the nodes cross
the high-density
$(0,\pi/a)$ areas {\it no matter the details of the
interaction.}

In conclusion, we have proposed an efficient computational
approach to the multiband gap equation.
Our selfconsistent calculations of the gap
in UPd$_2$Al$_3$ demonstrate that the only
accessible gap symmetry with nodes transforms according to 
the $A_{1g}$ irreducible representation of the 
$D_{6h}$ point group, independent of whether 
the pairing potential 
is dominated by small-$q$ attractive processes
involving phonons or by 
spin fluctuations. The robustness of the $A_{1g}$ representation
is analyzed
to be due to the     
presence of high-density regions 
of the relevant bands in the vicinity
of the $\Gamma$ and $\rm A$ points, that preclude node
formation at these points.
As a general rule we obtain that nodes must be 
as far as possible from high-density areas in the phase space.

We are grateful to H. Adrian, N. Bernhoeft,   
P. Fulde, M. Huth, M. Lang, P. Thalmeier, N. Sato and
F. Steglich for illuminating discussions. We thank M. Huth for the
figure of the irreducible representations of the $D_{6h}$ group.



\newpage
\begin{figure}[tbp]
\caption{
({\bf a}) The hexagonal Brillouin zone
with high symmetry points. In the AFM phase the BZ
is reduced by a factor two along the $z$ axis 
(denoted by the primed letters).
({\bf b}) The energy bands along high symmetry directions
in the paramagnetic, and ({\bf c}) the AFM phase of UPd$_2$Al$_3$.
The relevant bands for superconductivity
are shown by the dotted lines. 
}
\label{fig1}
\end{figure}

\begin{figure}[tbp]
\caption{
The even parity $d$-wave-type irreducible representations
of the $D_{6h}$ point group. While in principle all are accessible
for the gap, only the $A_{1g}$ results from our selfconsistent
gap calculations.
}
\label{fig2}
\end{figure}

\begin{figure}[tbp]
\caption{
Examples of the selfconsistent solutions of the gap equation
in characteristic planes 
(with sides indicated) containing the $\Gamma$ point (in the center)
and the $z$ axis.
The thick black lines
show the nodes and in the dark gray areas the gap is negative.
All nodes cut the $z$ axis as in the $A_{1g}$
representation. Shown are gaps computed  with the
small-$q$ pairing and paramagnetic energy bands       
({\bf a}), and the spin-fluctuations pairing and AFM bands, ({\bf b}) 
and (${\bf c}$). 
}
\label{fig3}
\end{figure}

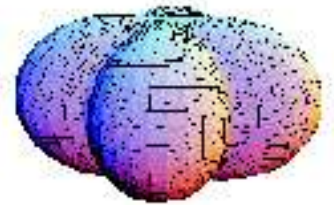
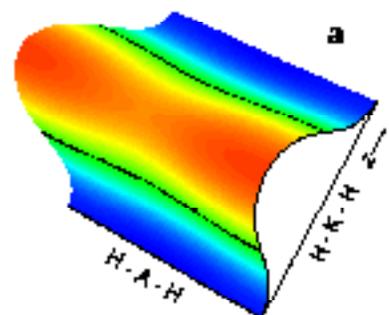
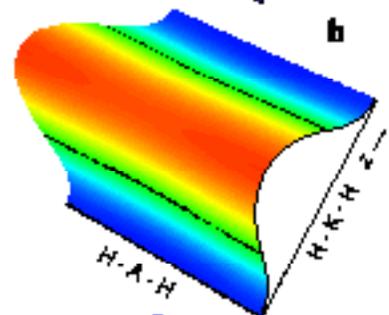
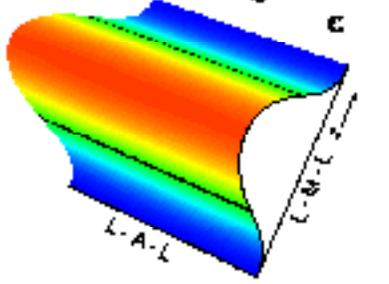
\end{document}